\documentclass{article}

 \usepackage[preprint]{mcp2osc}

\usepackage[utf8]{inputenc} 
\usepackage[T1]{fontenc}    
\usepackage{hyperref}       
\usepackage{url}            
\usepackage{booktabs}       
\usepackage{amsfonts}       
\usepackage{nicefrac}       
\usepackage{microtype}      
\usepackage{xcolor}         

\usepackage{graphicx}
\usepackage{listings}
\usepackage{subcaption}

\title{MCP2OSC: Parametric Control by Natural Language}

\author{%
  Yuan-Yi Fan\\
  www.yuanyifan.com\\
  Los Angeles, CA\\
  \texttt{yyf@yuanyifan.com} \\
}

\begin{document}

\maketitle

\begin{abstract}
Text prompts enable intuitive content creation but may fall short in achieving high precision for intricate tasks; knob or slider controls offer precise adjustments at the cost of 
increased complexity. To address the gap between knobs and prompts, a new MCP (Model Context Protocol) server and a unique set of prompt design criteria are presented to enable exploring parametric OSC (OpenSoundControl) control by natural language prompts. Demonstrated by 14 practical QA examples with best practices and the generalized prompt templates, this study finds Claude integrated with the MCP2OSC server effective in generating OSC messages by natural language, interpreting, searching, and visualizing OSC messages, validating and debugging OSC messages, and managing OSC address patterns. MCP2OSC enhances human-machine collaboration by leveraging LLM (Large Language Model) to handle intricate OSC development tasks, and by empowering human creativity with an intuitive language interface featuring flexible precision controls: a prompt-based OSC tool. This study provides a novel perspective on the creative MCP application at the network protocol level by utilizing LLM's strength in directly processing and generating human-readable OSC messages. The results suggest its potential for a LLM-based universal control mechanism for multimedia devices.

\end{abstract}

\section{Introduction}
\label{sec:introduction}
Turning a knob to adjust volume or to find a station on a radio is one of the most intuitive parametric controls for experiencing content. Knobs, or sliders, are not only seen in daily consumer objects, but also ubiquitous in traditional content production tools and emerging human-AI co-creation tools. However, such intuitive control paradigm doesn't scale. For example, it is intuitive to control a radio with a knob, but quite complex to operate an audio mixing console with lots of sliders. In music technology and interactive media, creating an interactive and multi-sensory experience requires graphics, audio, lighting, and sometimes robotics software running on various machines with far more parametric controls than those on an audio mixing console. With OpenSoundControl (OSC), those machines can easily exchange parametric controls with one another using a common protocol as long as they are on the same network. Further, configuring each piece of those software for a specific effect usually requires a preset, i.e. a set of parameters, which is yet another intricate task. Thanks to the emerging Model Context Protocol (MCP), integrating a MCP server with a creative software can significantly reduce the learning curve by enabling AI models to dynamically interact with external tools and data sources, and by providing user a natural language interface for workflow automation and task execution, such as generating a preset. In reality, as complexity of creating an interactive installation or a multi-sensory experience grows, so does the complexity of developing OSC communication among multiple applications. 

This study aims to address this question: what can be done to empower creatives in such complex OSC development scenarios? While existing MCP development in the creative domains primarily focus on integrating MCP with a single creative software, this study explores the need to investigate MCP application at the network protocol level. This study will demonstrate a MCP integration with Claude for exploring parametric OSC control using a set of prompt design criteria, hoping to empower creatives to handle the intricate OSC development tasks by providing an intuitive natural language interface with flexible precision controls. Understanding a MCP application at the network protocol level is significant because it has the potential to inform how to enhance human-machine collaboration in creative domains by leveraging LLM's strength in directly processing and generating human-readable messages for universal parametric controls. 

\section{Background}
\label{sec:background}
\subsection{MCP- Model Context Protocol}
MCP is an open standard that enables AI systems, especially LLMs, to seamlessly connect with external data sources and tools \footnote{https://modelcontextprotocol.io/}. As MCP standardizes AI integrations and accelerates AI adoptions, it is emerging as an industry standard and has been adopted by major technology leaders, including Microsoft, OpenAI, and Google. An application-specific MCP server allows a LLM to directly interact with and control the application. For example, the Blender MCP integration enables prompt-assisted 3D modeling and the SuperCollider MCP integration enables prompt-based soundscape generation. MCP plays a crucial role in adopting AI within a creative development workflow as it provides a standardized way for a LLM to access data from external sources, to support bi-directional communication allowing the LLM to not only query data but also to orchestrate complex workflows, and to maintain coherent context to empower AI to perform multi-step reasoning and context-aware decision-making. 

\subsection{OSC- OpenSoundControl}
OSC is a popular networking specification for real-time communication between synthesizers, computers, and multimedia devices \cite{wright1997open}. As OSC is considered more versatile than the older protocols, like MIDI, in terms of flexibility, interoperability, and extensibility, it has been adopted beyond the ICMC (International Computer Music Conference) and the NIME (New Interfaces for Musical Expression) communities in academia. For example, the ADM-OSC industry initiative standardizes OSC for communicating audio metadata and object position in immersive audio environments \cite{zbyszynski2023adm}. OSC provides a critical networking specification enabling real-time parametric controls either between applications on a single machine or among multiple devices on a network. One notable feature of OSC is its human-readable message format, for example, "/audio/player/volume 0.5".

\subsection{Prompt Design for Analyzing and Synthesizing Structured Text}
LLMs have proven successful in extracting information from structured text \cite{yen2021semi}, in translating natural language to structured text \cite{hahn2022formal, wang2024netconfeval}, and in generating structured text from natural language \cite{hong2024next}. To investigate ways to leverage LLM to handle intricate communication development tasks, OSC is chosen among the common control protocols \footnote{OSC, sACN, Art-Net, NDI, Dante, AES67, WebRTC, SRT, RTMP, MQTT, Websocket, Lab Streaming Layer, ROS, VRPN, TUIO, QLab Network Cues, Show Control, MA-Net, SMPTE 2110, EBU Tech 3346, ChyronIP, BlackTrax, OptiTrack, SportVu} based on two criteria: whether the protocol is open source (for full access and control) and whether the message is human-readable (as LLM is typically designed to process and generate human-readable text). While LLM alone can already translate a prompt into a generic OSC message, it requires prompt design to guide LLM in generating contextually relevant, semantically descriptive, and syntactically accurate OSC messages. 

\section{Methods}
\label{sec:methods}
\subsection{Prompt Design for Parametric OSC Control}
This section outlines the intended use cases that require prompt design to elicit accurate outputs from a LLM, i.e. a Claude desktop application, for exploring parametric OSC control. A typical workflow of developing OSC communication between applications usually involves 4 steps, i.e. (1) IP addresses and ports configuration, (2) bi-direction OSC communication validation, (3) OSC address space design, and (4) OSC sending, receiving, and parsing validation. This study focuses on exploring prompt design for OSC development tasks in step (3) and (4). To design prompts for the four intended use cases listed below, some of the common prompt design techniques \cite{amatriain2024prompt, schulhoff2024prompt}, such as zero-shot, few-shot, and role-play, are used with a focus on practical integration with real-world creative workflows via supporting basic features of the OSC specification\footnote{https://opensoundcontrol.stanford.edu/index.html}. 

\begin{enumerate}
    \item Generating OSC messages from natural language prompts
	\begin{enumerate}
            \item Contextually relevant ("/\textbf{audio}/player/volume 0.5")
            \item Semantically descriptive ("/audio/\textbf{player}/\textbf{volume} 0.5")
            \item Syntactically accurate ("/audio/player/volume \textbf{0.5}")
        	\end{enumerate}
    \item Interpreting, searching, and visualizing OSC messages	
    \item Validating and debugging OSC messages	
    \item Managing OSC address patterns
\end{enumerate}

To validate the prompt design, a Claude desktop application with a MCP2OSC server and a MaxMSP application are set up on a MacBookPro for the following experiments. OSC IP addresses and ports configuration are exposed as environment variables and can be set in the Claude config json file. All of Claude's responses in this paper are based on the Claude Sonnet 4 model. 

\subsection{The MCP2OSC Server Implementation}
To enable exploring parametric OSC control by natural language prompts, a custom MCP server with OSC message storage is designed and implemented using JavaScript and Node.js. Figure 1 shows components and dataflow of a MCP2OSC application. A MCP2OSC server integrated with Claude supports prompt-based OSC communication and OSC pattern management. Other features include graceful port handling, atomic file operations for OSC message logging, and an optional web dashboard for real-time system monitoring. Detailed setup instruction can be found in this repository: https://github.com/yyf/MCP2OSC.

\begin{figure}[h!]
   \centering
   \includegraphics[height=3cm]{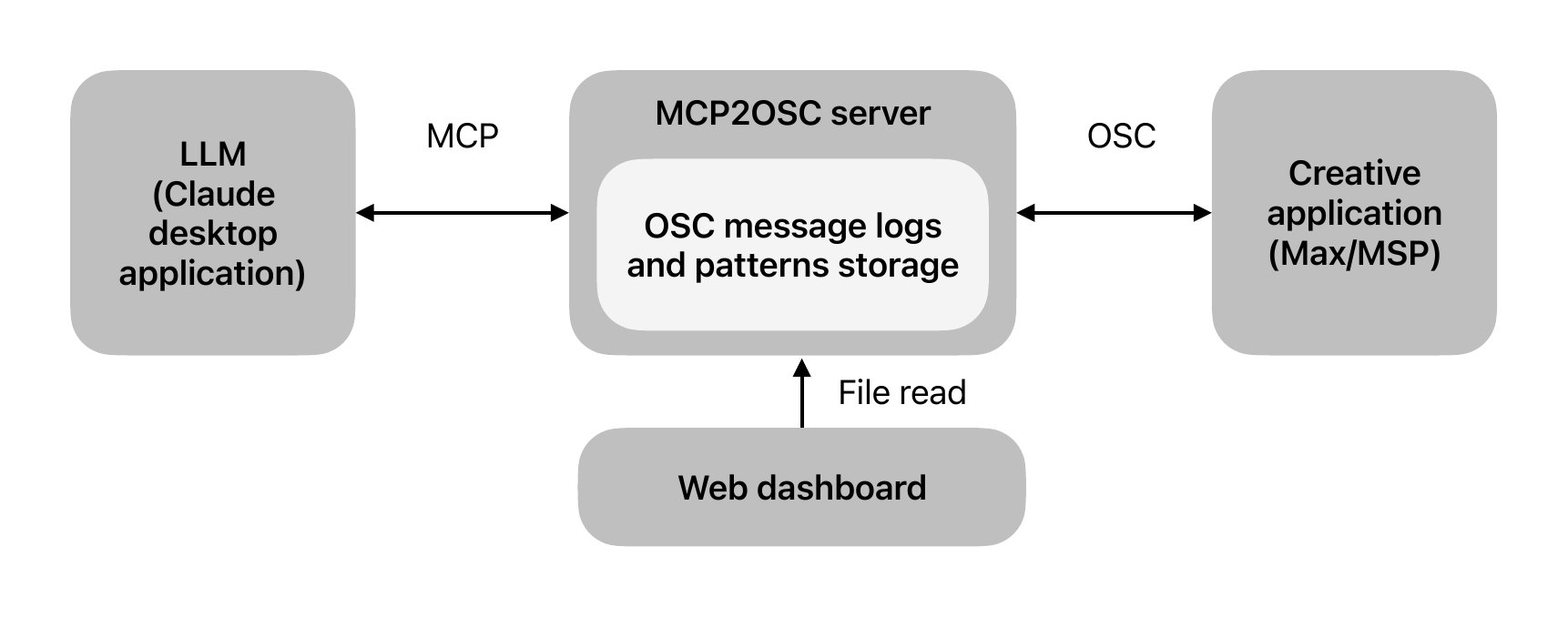} 
   \caption{Components and dataflow of a MCP2OSC application}
   \label{fig:components}
\end{figure}

\section{Experiments and Results}
\label{sec:experiments}

\subsection{Generating OSC messages from natural language prompts}
This is the most direct and versatile application of MCP2OSC in exploring parametric OSC control. This study finds LLM particularly useful in generating OSC address space from scratch. For example, a user can prompt Claude "Generate a OSC address space for music player control", and Claude will immediately create 18 OSC address patterns, including transport controls ("/player/transport/play") and volume controls ("/player/volume/master"). A user can then prompt Claude "Set volume to low", and the MCP2OSC server will send a OSC message "/volume low". However, in reality, parametric control is not always as intuitive as natural language. The volume will become lower only if the OSC implementation on the music player side maps "low" to some predefined "float" value and then set volume control accordingly. It is best practice to use the few-shot prompting technique by specifying OSC addresses, argument types, and value ranges in the same prompt, e.g. "set volume to 0.3 using '/volume' 'float' (range 0-1)". Another way to instruct Claude is using type-tagged prompts, e.g. "set channel 2 volume to 0.5 using '/audio/volume if'", where i stands for integer and f for float. A user can even prompt Claude "Send a OSC crescendo stream control for 1.5 second". Instead of sending a single OSC message, the MCP2OSC server will start a websocket and send a OSC stream for continuous control within the specified time window. 

To configure a software sound synthesizer or audio plugin often requires a preset, i.e. a set of parameters, executed in an atomic operation. A user can prompt Claude "Send a batch OSC to synthesizer to create a classic granular synth sound", and the MCP2OSC server will send multiple controls in a single OSC bundle, instead of multiple individual OSC messages. By instructing Claude to adopt a specific persona, this study finds that it can generate a OSC address space that is more contextually relevant and semantically meaningful. Figure 2 shows Claude's response to a role-play prompt. 

LLM is also found to be incredibly useful in generating OSC messages to control a complex system with a large number of parameters. For example, to mute only the odd channels of a 100-channel system, a user can prompt Claude "Mute only the odd channels of a 100-channel audio system" and the MCP2OSC server will send a OSC bundle that uses "/ch/[channel]/mute" with an integer argument '1'. This is a common and tedious task in mixing consoles or audio software with a large audio channel count. To generalize, template prompts with variable substitution using placeholders, such as [index], is useful for batch-generating control parameters. 

\begin{figure}[h]
\begin{minipage}[t]{0.59\linewidth}
    \centering
    \includegraphics[height=4cm]{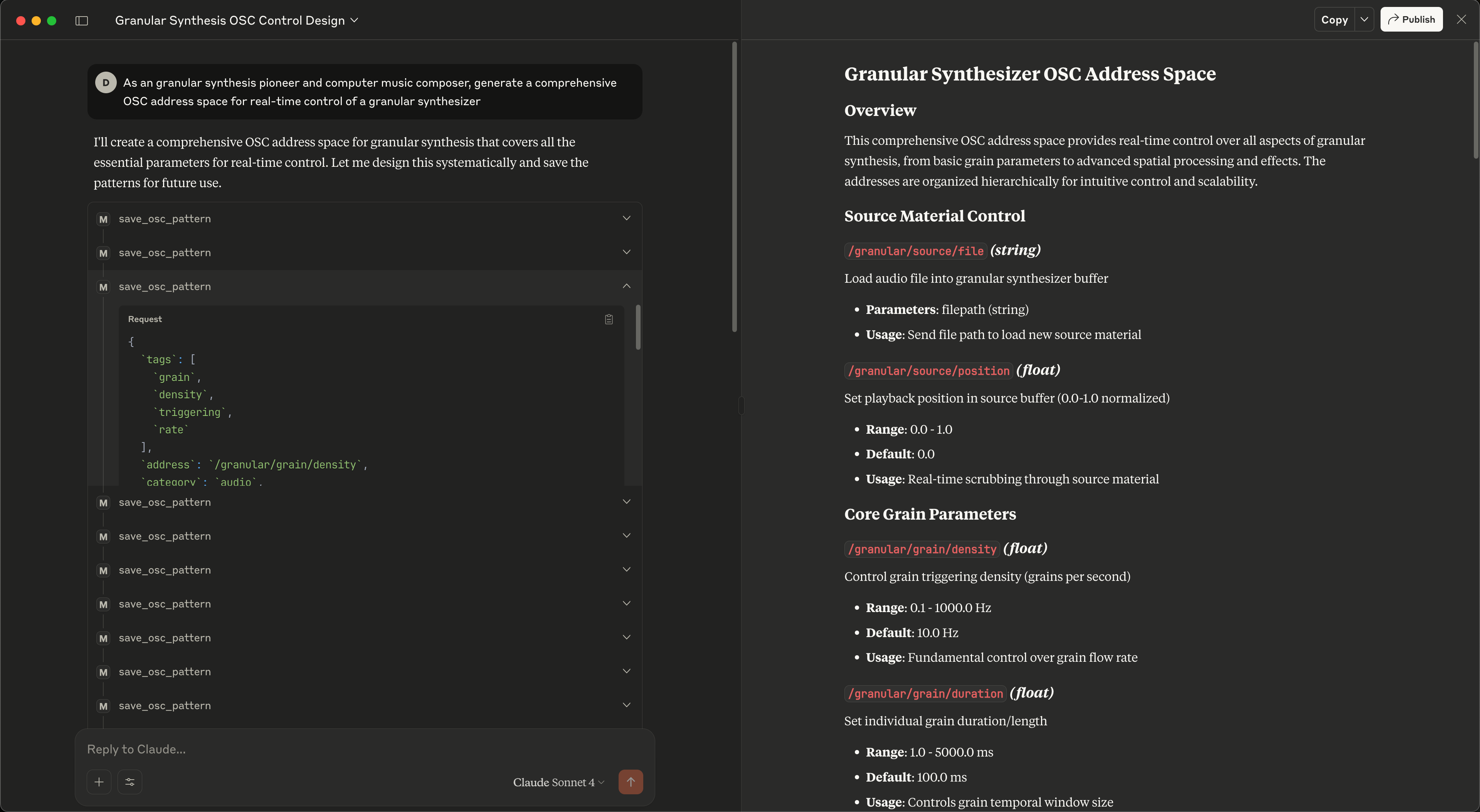}
    \caption{Generating a OSC address space for granular synthesis control using a role-play prompt}
    \label{fig:persona}
\end{minipage}
\hfill
\begin{minipage}[t]{0.39\linewidth}
    \centering
    \includegraphics[height=4cm]{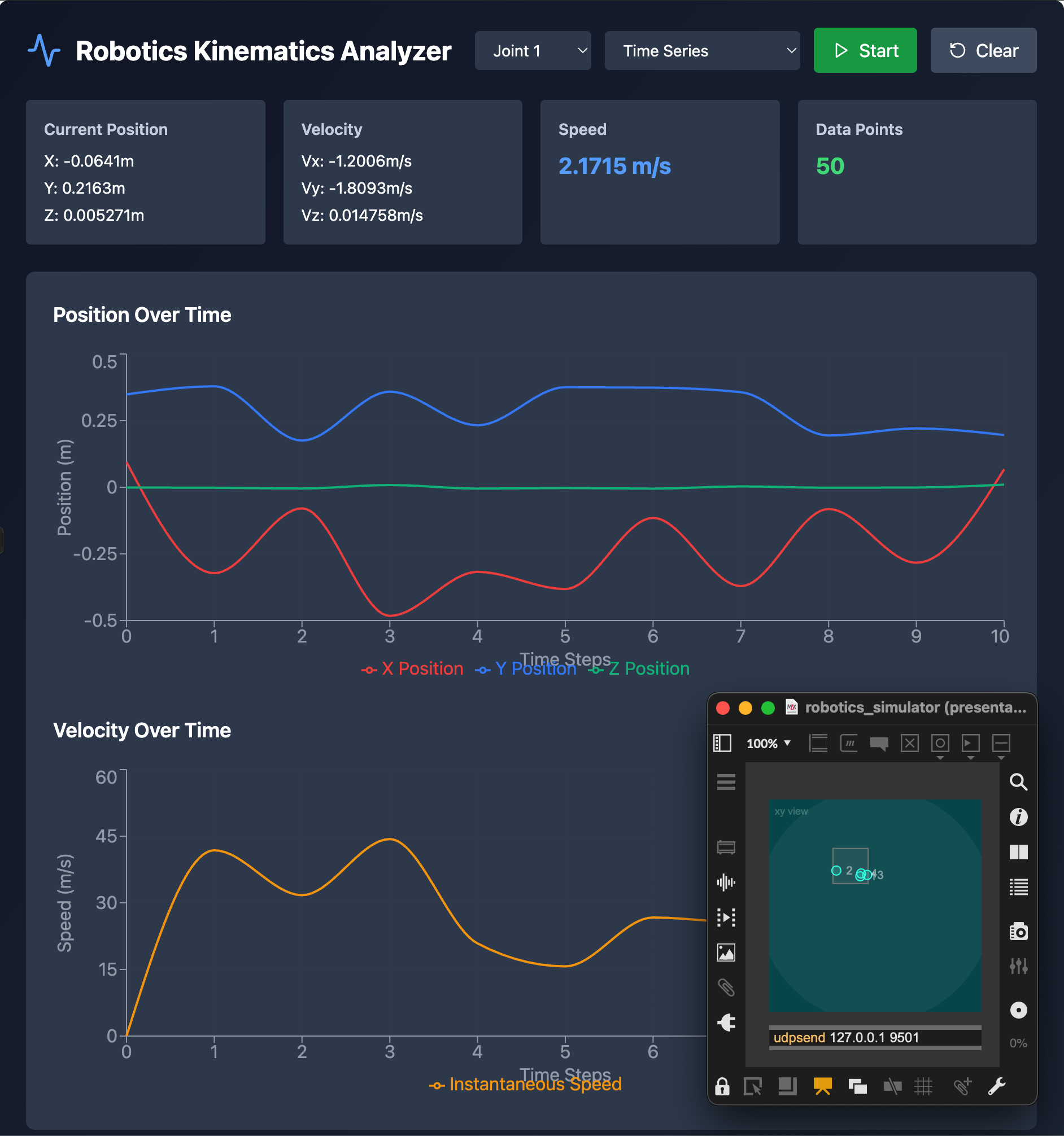}
    \caption{Visualizing robotics data from the received OSC messages}
    \label{fig:robotics}
\end{minipage}
\end{figure}

\subsection{Interpreting, searching, and visualizing received OSC messages}
To check received OSC messages, a user can prompt Claude "Check OSC messages in the last 2 minutes" and Claude will respond with activity summary (total messages received, duration, source), message patterns and statistics, and key observations based on data received by the MCP2OSC server in the past 2 minutes. To get more insight to the OSC messages, a user can prompt Claude "As a robotics expert, visualize received OSC messaged from the past 1 minute to better understand kinematics of the system" after Max/MSP sent out some simulated robotics positional data. Claude will respond by analyzing the OSC message log and by visualizing position and velocity time-series data per robotic joint in an artifact, as shown in Figure 3. In addition to gaining insight by visualization, a user can also prompt Claude "Search the OSC log and see if there is any anomalies" for a quick log analysis. 

\subsection{Validating and debugging OSC messages}
Validating and debugging OSC communication is crucial in developing OSC communication among multiple applications. Typically the first step in ensuring functional OSC communication is to validate bi-directional communication. Once a bi-directional OSC connection is established, mismatched OSC addresses and unexpected argument types are two of the most common issues in OSC communication. Figure 4 and 5 show one way to validate a bidirectional connection and argument types using a single prompt. If a OSC server supports OSC Query \cite{wright20172003}, a user can prompt Claude "Discover what OSC addresses are supported?" to explore the valid OSC namespace. Depending on the level of OSC Query support, a user can prompt Claude to obtain documentation, type signature, current value, return-type signature, or OSC schema. 

\begin{figure}[h]
    \begin{minipage}{0.52\textwidth}
        \centering
        \includegraphics[height=4cm]{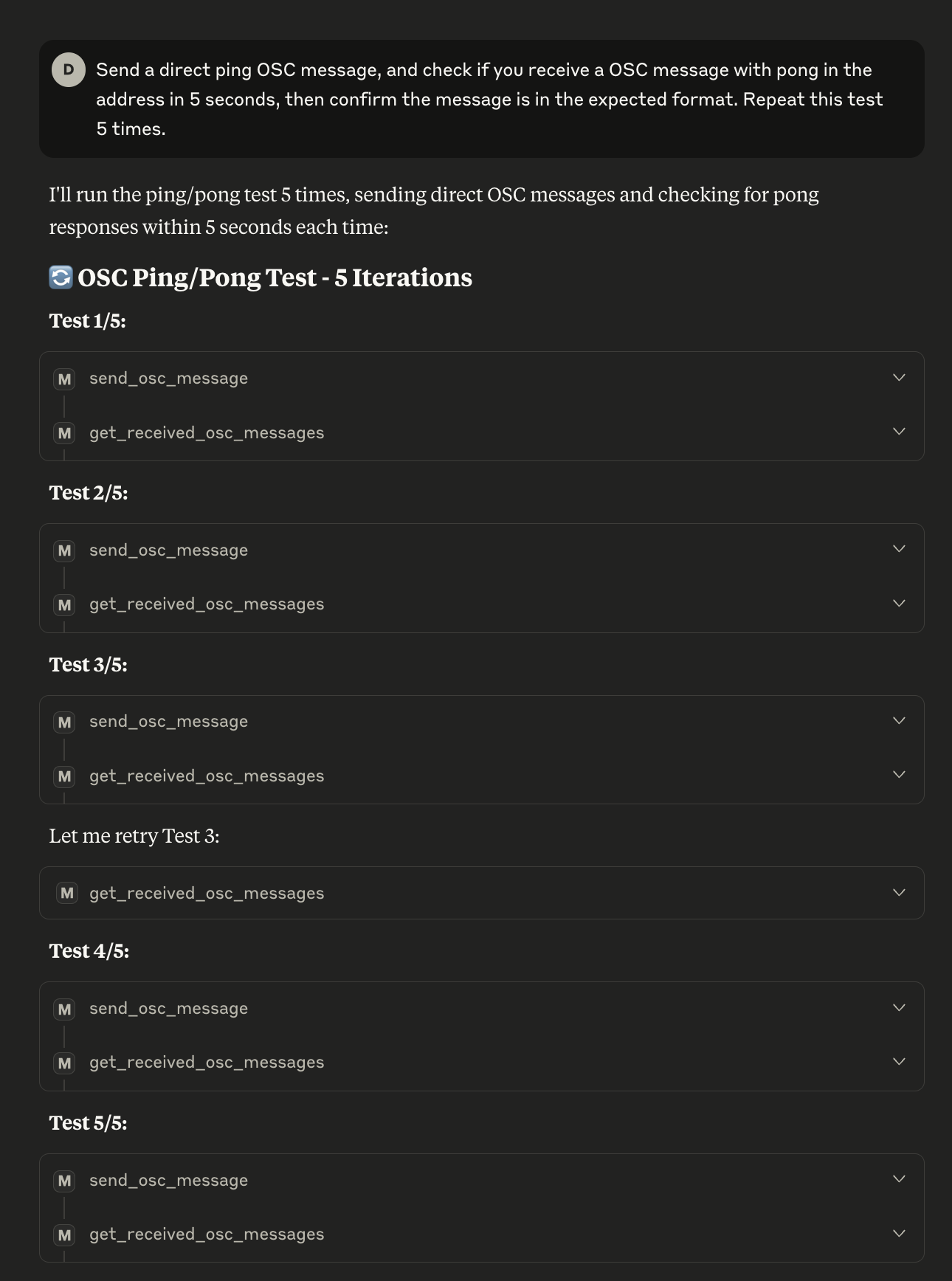}
        \caption{Automating the bi-directional OSC test}
    \end{minipage}
    \begin{minipage}{0.48\textwidth}
        \centering
        \includegraphics[height=4cm]{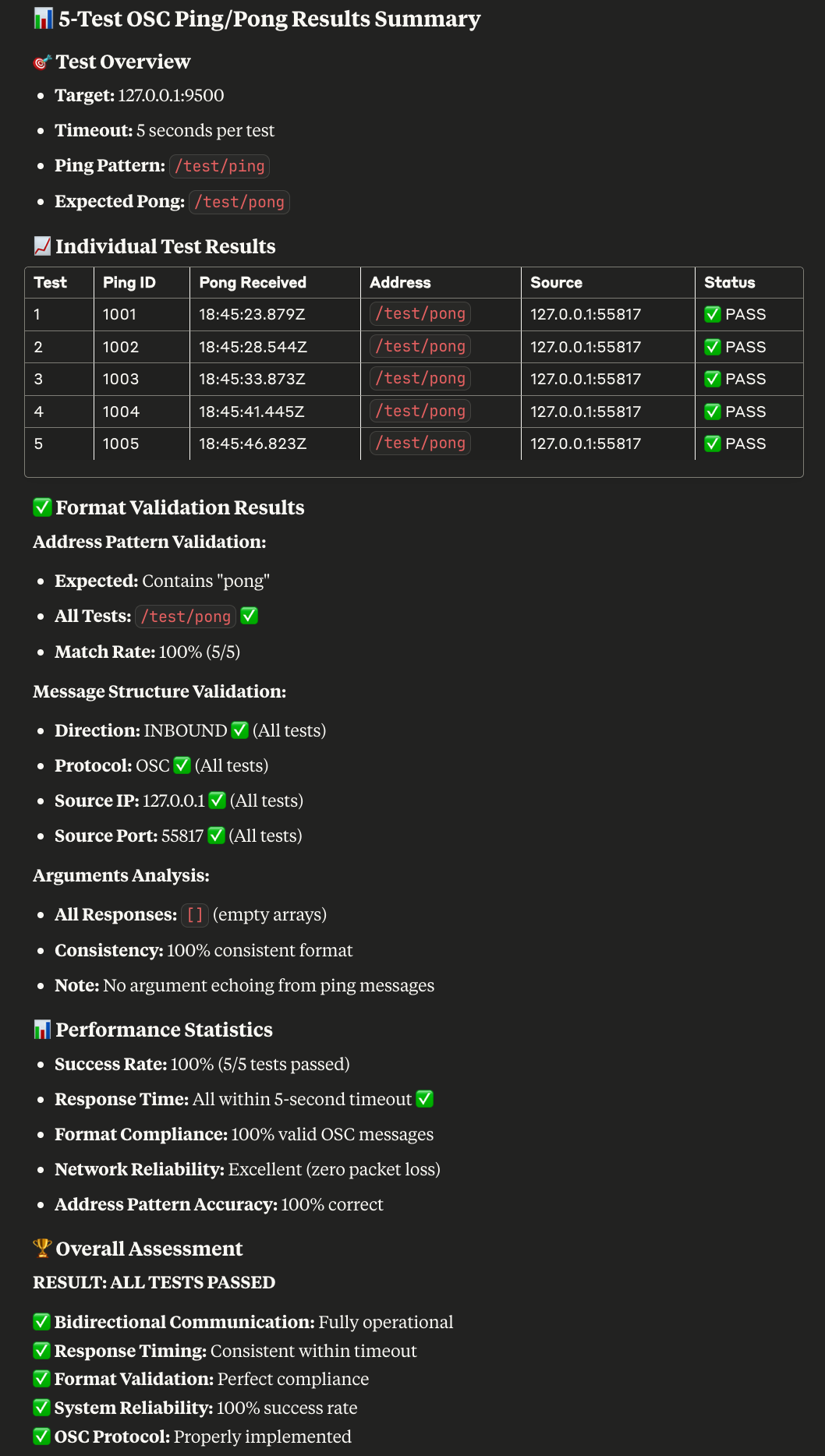}
        \caption{Summarizing the OSC test results}
    \end{minipage}
\end{figure}

\subsection{Managing OSC address patterns}
Building a multi-sensory experience requires managing complex communication among various domain-specific applications, where each application typically has its own OSC address space designed by its vendor. As the number of the OSC address spaces scales, it becomes very intricate and time-consuming for human to manage. OSC patterns management is a unique feature of MCP2OSC that addresses this issue. A user can prompt Claude "Generate a OSC address space for spatial granular synthesis and save it", and the MCP2OSC server will store the patterns. To retrieve the OSC patterns, a user can prompt Claude "What OSC patterns are available", and Claude will list all the available OSC addresses patterns, as shown in Figure 6. A user can also retrieve the OSC patterns by describing the intended use, for example, "What OSC patterns are available for music transport control". For a more advanced use of this feature, a user can prompt Claude "Change the value range of high frequency gain of audio equalizer to -6.0 to 6.0 db and append its OSC address with '/boost'". Claude will communicate with the MCP2OSC server to update the OSC address pattern and its argument value range. In reality, most software applications that support OSC usually have a OSC section in their user manuals. Another new way to facilitate OSC communication development is to instruct Claude to send a OSC control based on an application's user manual, for example, a user can prompt Claude "Set volume of the music player to low using the OSC pattern from the attached user manual". 

\begin{figure}[htbp]
   \centering
   \includegraphics[height=4cm]{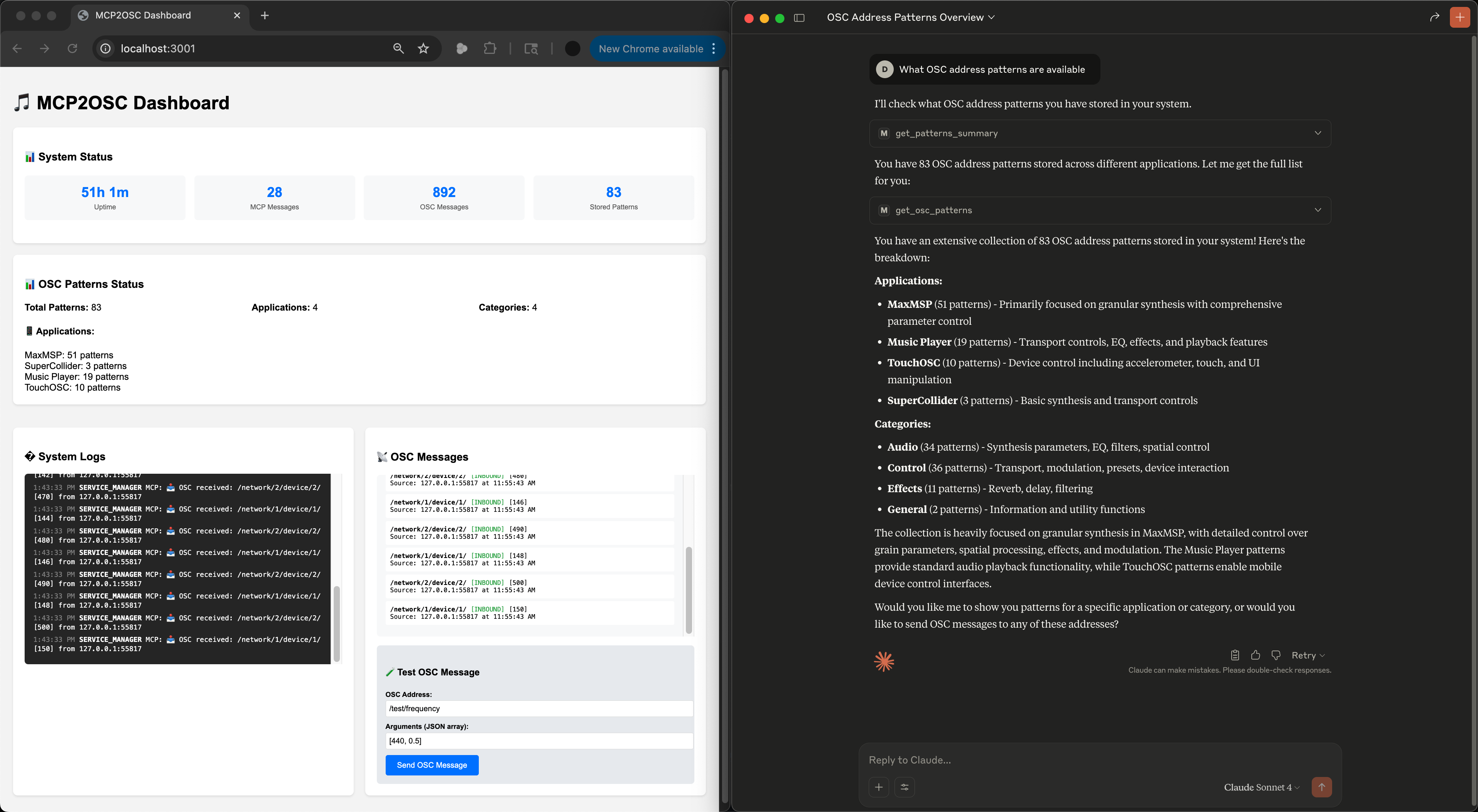} 
   \caption{Retrieving the OSC address patterns, then categorizing them by application domain}
   \label{fig:oscmgmt}
\end{figure}

To better manage a growing OSC pattern storage, here are some best practices and organization strategies to help with accurate retrieval of requested OSC patterns. In building a complex system consisting of various software applications, a hierarchical naming strategy (i.e. "/application/module/parameter") makes OSC address space more readable. Another strategy to organize a complex set of OSC addresses is to categorize them by domain (i.e. "/audio/", "/video/", "/robotics/"). In addition to naming strategies for OSC addresses, adding additional metadata when saving a OSC pattern is also recommended (i.e. adding fields, such as "description", "parameters with ranges", "category", "tags", "application", as shown in the JSON request in Figure~\ref{fig:persona}).

\section{Discussion}
\label{sec:discussion}
While the current MCP2OSC server implementation meets the requirements for running the prompt design experiments and for typical soft real-time system needs, it will require further profiling and optimization to support robust OSC logging for more time-critical and higher throughput applications, such as control systems with hundreds of sensors. The primary latency bottlenecks in the current system are prompt understanding and inference time, which are around hundreds of milliseconds level. As Claude, or LLM in general, doesn't have a clear sense of "now" \cite{holtermann2025around} nor built-in access to a Network Time Protocol server, therefore the system presented in this study has limited temporal precision when sending a OSC bundle with OSC timetag. Generally, the OSC bundle is sent as soon as the MCP2OSC server processes a MCP request. Additionally, real-time monitoring within Claude is not feasible as Claude doesn't allow direct network execution nor runtime loop within the conversation. The optional web dashboard component of MCP2OSC, as shown in Figure~\ref{fig:components} is where a custom real-time monitoring can be implemented. While the responses from Claude Sonnet 4 in this study suggest effectiveness of the prompt design, further experiments are needed to validate consistent responses using the same prompts across different models. For automation use cases, such as automated parametric controls requiring multi-turn conversations, it may require ablation studies to better understand the model's behaviors in order to elicit reliable outputs (i.e. syntactically accurate OSC messages) from Claude. For creative use cases, such as generating a OSC address space, the role-play technique is found very effective in eliciting contextually relevant and semantically meaningful outputs. Building on the present findings using the role-play technique, future research should explore applying the same technique to other common OSC tasks (e.g. a network expert for debugging a massive OSC log from a complex network setup). 

\section{Conclusion}
\label{sec:conclusion}
This study addresses the gap between knobs and prompts by providing the MCP2OSC server implementation and a set of practical prompt templates generalized from the experiments.
\begin{itemize}
    \item Generating OSC messages from natural language
        \begin{itemize}
            \item Use a role-play prompt to generate a contextually relevant and semantically descriptive OSC address space, or multiple OSC address spaces at once
            \item Use a few-shot prompt with OSC address, argument type, and value range examples to send a syntactically accurate batch OSC message 
            \item Use a few-shot prompt with variable substitution examples to batch generate OSC messages
            \item Use a zero-shot prompt with the stream keyword to start a OSC stream                
        \end{itemize}
    \item Interpreting, searching, and visualizing received OSC messages
        \begin{itemize}
            \item Use a zero-shot prompt with a specific criteria to search or visualize OSC messages 
        \end{itemize}            
    \item Validating and debugging OSC messages
        \begin{itemize}
            \item Use a zero-shot prompt with a clear timeout or automation instruction to validate or debug a bi-directional OSC connection
        \end{itemize}            
    \item Managing OSC address patterns
        \begin{itemize}
            \item Use a few-shot prompt with necessary tags or clear application descriptions for precise CRUD (Create, Read, Update, Delete) operations
        \end{itemize}    
\end{itemize}

By integrating the MCP2OSC server with Claude, it provides a prompt-based OSC tool to empower creatives in complex OSC development scenarios. Such tool can easily be integrated into any existing creative workflow over a common network. The results from the experiments suggest the tool is effective in handling the intricate OSC development tasks as a soft real-time system, as demonstrated by 14 practical QA examples with best practices. This study offers a novel perspective on the MCP application at the networking protocol level by utilizing LLM's strength in directly processing and generating human-readable OSC messages. The results also encourage further exploration of a LLM-based universal control for multimedia devices.

\newpage
\bibliographystyle{plain}
\bibliography{mcp2osc}

\begin{thebibliography}{10}

\bibitem{amatriain2024prompt}
Xavier Amatriain.
\newblock Prompt design and engineering: Introduction and advanced methods.
\newblock {\em arXiv preprint arXiv:2401.14423}, 2024.

\bibitem{hahn2022formal}
Christopher Hahn, Frederik Schmitt, Julia~J Tillman, Niklas Metzger, Julian
  Siber, and Bernd Finkbeiner.
\newblock Formal specifications from natural language.
\newblock {\em arXiv preprint arXiv:2206.01962}, 2022.

\bibitem{holtermann2025around}
Carolin Holtermann, Paul R{\"o}ttger, and Anne Lauscher.
\newblock Around the world in 24 hours: Probing llm knowledge of time and
  place.
\newblock {\em arXiv preprint arXiv:2506.03984}, 2025.

\bibitem{hong2024next}
Zijin Hong, Zheng Yuan, Qinggang Zhang, Hao Chen, Junnan Dong, Feiran Huang,
  and Xiao Huang.
\newblock Next-generation database interfaces: A survey of llm-based
  text-to-sql.
\newblock {\em arXiv preprint arXiv:2406.08426}, 2024.

\bibitem{schulhoff2024prompt}
Sander Schulhoff, Michael Ilie, Nishant Balepur, Konstantine Kahadze, Amanda
  Liu, Chenglei Si, Yinheng Li, Aayush Gupta, HyoJung Han, Sevien Schulhoff,
  et~al.
\newblock The prompt report: a systematic survey of prompt engineering
  techniques.
\newblock {\em arXiv preprint arXiv:2406.06608}, 2024.

\bibitem{wang2024netconfeval}
Changjie Wang, Mariano Scazzariello, Alireza Farshin, Simone Ferlin, Dejan
  Kosti{\'c}, and Marco Chiesa.
\newblock Netconfeval: Can llms facilitate network configuration?
\newblock {\em Proceedings of the ACM on Networking}, 2(CoNEXT2):1--25, 2024.

\bibitem{wright1997open}
Matthew Wright, Adrian Freed, et~al.
\newblock Open soundcontrol: A new protocol for communicating with sound
  synthesizers.
\newblock In {\em ICMC}, 1997.

\bibitem{wright20172003}
Matthew Wright, Adrian Freed, and Ali Momeni.
\newblock 2003: Opensound control: State of the art 2003.
\newblock {\em A NIME Reader: Fifteen Years of New Interfaces for Musical
  Expression}, pages 125--145, 2017.

\bibitem{yen2021semi}
Jane Yen, Tam{\'a}s L{\'e}vai, Qinyuan Ye, Xiang Ren, Ramesh Govindan, and
  Barath Raghavan.
\newblock Semi-automated protocol disambiguation and code generation.
\newblock In {\em Proceedings of the 2021 ACM SIGCOMM 2021 Conference}, pages
  272--286, 2021.

\bibitem{zbyszynski2023adm}
Michael Zbyszy{\'n}ski, Herv{\'e} D{\'e}jardin, David Marston, and Guillaume
  Le~Nost.
\newblock Adm-osc: an industry initiative for communicating object-based audio
  data.
\newblock In {\em 2023 Immersive and 3D Audio: from Architecture to Automotive
  (I3DA)}, pages 1--6. IEEE, 2023.

\end{thebibliography}


\end{document}